\documentstyle[preprint,eqsecnum,epsfig,prd,aps,floats]{revtex} 

\begin{document}

\newcommand{\nl}{\nonumber \\}
\newcommand{\eq}[1]{Eq.~(\ref{#1})}

\preprint{\vbox{ \tighten {
		\hbox{MIT-CTP-2930}
		\hbox{PUPT-1903}
                \hbox{NSF-ITP-149}
		}  }}  
  
\title{Four-dimensional gravity on a thick domain wall}
  
\author{ Martin Gremm\thanks{email: gremm@feynman.princeton.edu }\thanks{ 
On leave of absence from MIT, Cambridge, MA 02139}} 
  
\address{Joseph Henry Laboratories, Princeton University, Princeton, NJ 08544\\
and\\ 
Institute for Theoretical Physics, University of California,
Santa Barbara, CA 93106}
 
\maketitle  
 
\begin{abstract} 
\tighten{ 
We consider an especially simple version of a thick domain wall in $AdS$ space
and investigate how four-dimensional gravity arises in this context. 
The model we consider has the advantage, that the equivalent quantum mechanics
problem can be stated in closed form. The potential in this Schr\"odinger
equation suggests that there could be resonances in the spectrum of the
continuum modes. We demonstrate that there are no such resonances in the
model we consider.
} 
\end{abstract} 
 
\newpage

\section{Introduction} 
 
Recently it was discovered that four-dimensional gravity can be realized
on a domain wall in five-dimensional space-time \cite{RS} (For a review
on domain walls see \cite{review}). The concrete example
studied in Ref.~\cite{RS} involves two regions of $AdS_5$ glued together
over a four-dimensional Minkowski slice. The fluctuations of the gravitational
field include a single normalizable zero mode, which gives rise to 
four-dimensional gravity on the domain wall. Since the ambient space is
non-compact, there is a continuous spectrum of massive modes. Normally this
results
in five-dimensional gravity being restored, but if the ambient space is a
suitable slice of
$AdS$, this does not happen. The couplings of the extra modes to matter on
the domain wall are sufficiently suppressed that integrating over all of
them still gives a subleading contribution to the gravitational interaction
between test masses on the domain wall. The original proposal was formulated
in pure five-dimensional gravity.  Various extensions to 
domain walls in gravity coupled to scalars \cite{mark,gubs}, time
dependent cosmological scenarios \cite{cosm}, higher dimensional
embeddings \cite{csaki}, and models with mass gaps for the continuum modes
\cite{bs,sn} have appeared in the literature. Several supergravity solutions
related to the model of \cite{RS} are known \cite{sdw}, but the original model
does not have a supersymmetric extension \cite{nogo}.

In this note we consider a solution of five-dimensional gravity 
coupled to scalars, which also does not have a supersymmetric
extension\footnote{We thank A.~Linde and R.~Kallosh for pointing this out to
us.}.  Our solution can be interpreted as a thick domain
wall interpolating between two asymptotic $AdS_5$ spaces, which makes it
a non-singular version of the setup in \cite{RS}. It is interesting to 
investigate how four-dimensional gravity arises on non-singular domain walls,
since they are a more realistic implementation of the scenario in
Ref.~\cite{RS}.

Thick domain walls are easy to construct, but all explicit solutions we are
aware of are too complicated for analytical computations. Our solution is
simple enough that the equivalent quantum mechanics
problem can be given in closed form. 
The potential in the equivalent quantum mechanics problem has
the form of a shallow well separated from the asymptotic region by a thick
potential barrier. A natural question is whether there are resonances (or
quasi-stationary states) in such potentials. From the point of view of a
four-dimensional observer, these states would give a quasi-discrete spectrum of
low mass KK states with unsuppressed couplings to matter on the domain wall.
This should be contrasted with the couplings of 
non-resonant modes which are small since they have to tunnel
through the potential barrier. If such resonant states exist, they change the
physics a four-dimensional observer sees. In order to address
questions of this type it is convenient to have a closed form expression 
for the quantum mechanics problem.  In this note we discuss one such example
and find that there are no resonances of the type described above. Thus we
conclude that the mechanism for localizing gravity on a thick domain wall is
exactly the same as in the thin wall case of \cite{RS}.

In section II we give a quick review of gravity coupled to scalars and outline
a method for finding solutions following \cite{gubs}. We use it to find a
particularly simple thick domain wall solution.
In section III we study fluctuations of the gravitational field
around our solution and describe how bulk modes interact with matter on
the domain wall.

\section{Gravity coupled to scalars}

In order to ensure four-dimensional Poincare invariance we assume that the
metric takes the form

\begin{equation}
\label{metric}
ds^2 = e^{2A(r)}\left( dx_0^2 - \sum_{i=1}^3 dx_i^2 \right) - dr^2.
\end{equation}
The action for five-dimensional gravity coupled to a single real scalar reads
\begin{equation}
S = \int d^4x dr \sqrt{g}\left( -\frac{1}{4} R +
\frac{1}{2}(\partial\phi)^2 - V(\phi) \right),
\end{equation}
and the equations of motion following from this action are
\begin{eqnarray}
\phi^{\prime\prime}+4A^\prime \phi^\prime &=&
\frac{\partial V(\phi)}{\partial\phi} \\
A^{\prime\prime} &=& -\frac{2}{3} \phi^{\prime 2} \\
A^{\prime 2} &=& -\frac{1}{3}V(\phi)+\frac{1}{6}\phi^{\prime 2}. \nonumber
\end{eqnarray}
The prime denotes differentiation with respect to $r$, and we have assumed that
both $\phi$ and $A$ are functions of $r$ only. 
If the potential for the scalar is given by 
\begin{equation}
V(\phi) = \frac{1}{8} \left( \frac{\partial W(\phi)}{\partial \phi}\right)^2
-\frac{1}{3} W(\phi)^2,
\end{equation}
setting 
\begin{equation}
\phi^\prime = \frac{1}{2}\frac{\partial W(\phi)}{\partial\phi}, \qquad 
A^\prime = -\frac{1}{3} W(\phi)
\end{equation}
yields a solution. 
This very useful first order formalism for obtaining solutions to the
equations of motion appeared first in the study of supergravity 
domain walls \cite{fo1} and was generalized in \cite{fo2,gubs} 
to include non-supersymmetric domain walls in various dimensions. 

To study domain wall solutions we have to choose a scalar potential with
several minima. A domain wall solution is characterized by a function $\phi(r)$
that asymptotes to different minima of the potential as $r\to \pm \infty$. 
In general $\phi(r)$ is smooth and contains a length scale that corresponds
to the thickness of the wall. 

It is straightforward to find such solution to the equations of motion. Some
examples are discussed in \cite{fo1,gubs} and many others can be constructed
along
the same lines. However most of these solutions yield equations that are too
complicated for an analytical treatment in closed form. In this note we
present a very specific example of a thick $AdS$ domain wall, where most of the 
calculations can be done analytically. 

We choose a superpotential
\begin{equation}
W(\phi) = 3 b c \sin\left( \sqrt{ \frac{2}{3b}}\phi\right),
\end{equation}
which gives rise to 
\begin{eqnarray}
V(\phi) &=& \frac{3 bc^2}{8}
\left( (1-4b)-(1+4b)\cos\left(\sqrt{\frac{8}{3b}}\phi\right)\right), \\
\label{ar}
A(r) &=& -b \ln\left(2 \cosh(c r) \right), \\
\phi(r) &=& \sqrt{6b}\, {\rm arctan}
	\left( \tanh\left( \frac{c r}{2}\right)\right),
\end{eqnarray}
where we have set an integration constant that corresponds to a shift in $A(r)$
to zero.
For $r\to\pm\infty$ we have $A(r) \sim - bc|r|$, so the metric, \eq{metric}, 
reduces to $AdS$ far from the domain wall at $r=0$. Generic $AdS$ domain wall
solutions have two free parameters, one for the asymptotic $AdS$ curvature
and another for the width of the wall. In the solution above the $AdS$
curvature is given by $b c$ and the thickness of the wall is parametrized by
$c$.

\section{Metric fluctuations}

In gravity coupled to scalars one cannot discuss fluctuations of the 
metric around the background given in the previous section without including
fluctuations of the scalar as well. The general treatment of these
fluctuations is rather complicated, since one has to solve a complicated
system of coupled differential equations. However, there is a sector of the 
metric fluctuations that decouples from the scalars \cite{gubs}, and these
fluctuations can be treated analytically. 

For the metric fluctuations we adopt a 
gauge such that the perturbed metric takes the form
\begin{equation}
ds^2 = e^{2A(r)} (\eta_{ij} + h_{ij}) dx^idx^j - dr^2,
\end{equation}
where $i=0,...,3$ and $\eta_{ij} = {\rm diag}(1,-1,-1,-1)$.
It is straightforward but tedious to obtain the coupled
equations of motion for the scalar fluctuation, $\tilde{\phi}$,
and $h_{ij}$ \cite{gubs}. In order to prove stability of the solution given
in the previous section, we would have to show that there are no negative mass
solutions to these equations. Unfortunately, this is a rather daunting task,
and we will not attempt it here. 

The transverse and traceless part of the metric fluctuation,
$\bar h_{ij}$, 
decouples from the scalar and satisfies a much simpler equation of motion
\cite{gubs}
\begin{equation}
\label{weq}
\left( \partial_r^2 + 4A^\prime\partial_r-e^{-2A}\Box \right) \bar h_{ij} = 0,
\end{equation}
where $\Box$ is the four-dimensional wave operator. In \cite{RS} it proved
useful to recast this equation in a form similar to Schr\"odinger's equation.
To that end we change coordinates to $z = \int e^{-A(r)} dr$.
The metric takes the form
\begin{equation}
ds^2 = e^{2A(z)}\left( dx_0^2 - \sum_i dx_i^2 - dz^2 \right).
\end{equation}
and the wave equation for the transverse traceless parts of $h_{ij}$ reads
\begin{equation}
\left( \partial_z^2+3A^\prime(z)\partial_z + \Box \right) \bar h_{ij} = 0.
\end{equation}
Making the ansatz $\bar h_{ij}= e^{ik\cdot x} e^{-3A/2} \psi_{ij}(z)$, this
equation simplifies further to
\begin{equation}\label{se}
\left( -\partial_z^2 + V_{QM}(z) - k^2 \right) \psi(z) = 0,
\end{equation}
where we have dropped the indices on $\psi(z)$ and introduced the potential
$V_{QM} = \frac{9}{4} A^\prime(z)^2 + \frac{3}{2} A^{\prime\prime}(z)$.

Using \eq{ar} and the definition of the new variable $z$ we find
$z= \int dr 2^b \cosh^b(c r)$.  For integer $b$ these integrals are easy to
do, but if $b$ is even the inversion is never possible in closed form and for
odd $b$ inverting the expression for $z$ requires solving a degree $b$
polynomial equation. In the following we will set $b=1$. The $b=3$ case is
also tractable, but the resulting equations are much more complicated and
the qualitative behavior of the solution is the same as in the $b=1$ case.

Note that setting $b=1$ leaves only one 
free parameter, $c$, which controls the $AdS$ curvature and the thickness of
the domain wall. With $b=1$, we cannot take the thin wall limit without
sending the
$AdS$ curvature to infinity at the same time. Since we are interested in
studying thick walls, this is not a serious handicap.

For the $b=1$ solution we have $A(z) = -\ln( \sqrt{4 + c^2 z^2} )$,
which gives
\begin{equation}\label{pot}
V_{QM} = \frac{3 c^2}{4} \frac{(-8 + 5 c^2 z^2)}{(4+c^2z^2)^2}.
\end{equation}

The spectrum of eigenvalues, $k^2$, give the spectrum of graviton masses a
four dimensional observer at (or near) $z=0$ sees. In order to have a
four dimensional graviton as in \cite{RS} the lowest eigenfunction of \eq{se}
should have eigenvalue $k^2=0$. We find one normalizable eigenfunction given
by $\psi_0 = N/(4+c^2z^2)^\frac{3}{4}$, where $N$ is a normalization constant.
$\psi_0$ is the lowest energy eigenfunction, because it has no zeros. Thus
there is no instability from transverse traceless modes with $k^2<0$.

Since the potential vanishes for large $z$, this is the only
bound state. The wave functions for $k^2>0$ become plane waves at infinity. 
We were not able to find exact solutions of \eq{se} for $k^2>0$, but
qualitatively the process of localizing gravity on the domain wall is quite
clear.

\begin{figure}
\centerline{\epsfxsize = 8truecm \epsfbox{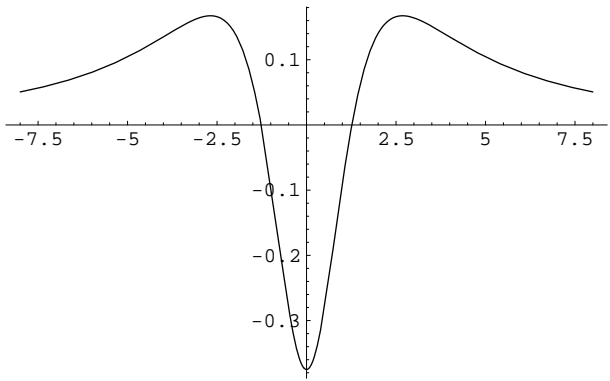} }
\tighten{
\caption{ The potential $V_{QM}$.}
\label{fig1}
}
\end{figure}

As in \cite{RS} any $k^2>0$ is allowed, so the masses of the extra modes are
continuous from zero. Normally this indicates that gravity is actually
five-dimensional, but for the thin $AdS$ domain wall the couplings of these 
modes to matter on the domain wall are suppressed enough that they do not
spoil four-dimensional gravity. We now discuss this somewhat mysterious
behavior in our non-singular setup. Let $V_{max}\sim c^2$ be the maximum of the
potential shown in Fig.~\ref{fig1}.
For modes with energies (masses) $k^2 \gg V_{max}$, the
potential is a small perturbation. These modes couple to matter on the domain
wall with regular strength, but since they are heavy, they yield subleading 
corrections to four-dimensional gravity mediated by the zero mode. 

\begin{figure}
\centerline{\epsfxsize = 8truecm \epsfbox{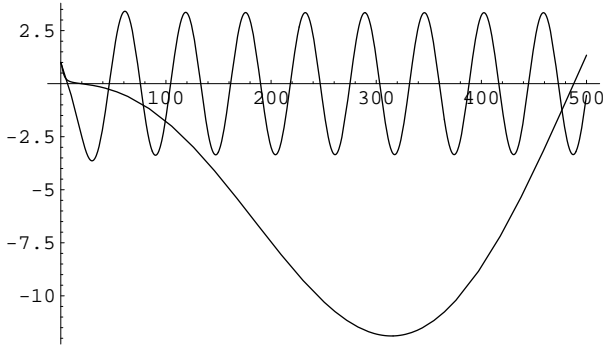} }
\tighten{
\caption{ Wave functions for moderate $k^2$ (rapidly oscillating) and small
$k^2$ (half period). The wave functions are shown in the unphysical
normalization $\psi(0)=1$.}
\label{fig2}
}
\end{figure}

Modes with $k^2 \le V_{max}$ could potentially exhibit a resonance structure.
Such resonant modes would have a disproportionately large amplitude in the
interior of the domain wall, while the amplitude of non-resonant modes should 
be much lager outside than in the interior. Since we do not have a solution 
for the continuum modes in closed form, we are forced to investigate this
question numerically. This analysis is simplified greatly by having a closed
expression for the potential \eq{pot}. 

Numerically integrating \eq{se} for various values of $k^2$ we find no
evidence for a resonance structure in the spectrum of the continuum modes.
Fig.~\ref{fig2} shows a mode with moderate $k^2$ and one with very low $k^2$.
Modes with intermediate values of $k^2$ smoothly interpolate between the two
solutions shown.
We have adopted the unphysical normalization, $\psi(0)=1$. To convert to the
physical normalization,
we use plane wave normalization for the wave functions at large
$z$. With this normalization the wave function for small $k^2$ 
is strongly suppressed at the origin. 
Since the coupling of a mode with mass $k^2$ to matter
located at or near $z=0$ is given by the amplitude of the properly normalized
wave function there, the modes with small $k^2$ are essentially
decoupled from the physics on the domain wall. Furthermore, the wave function
at the origin decreases monotonically as we lower $k^2$. Just as in the 
thin wall case of \cite{RS}, there is no evidence for a resonance structure in
the continuum modes. This is true both for the $b=1$ and the $b=3$ solution,
but it is not clear if this is a generic feature of thick $AdS$ domain walls,
or a special property of our simple examples. In any event we conclude that
there are thick domain walls on which the four-dimensional effective theory 
has a spectrum that is qualitatively very similar to the thin wall case. 

It is interesting to see how our solution reduces to the thin wall solution 
of \cite{RS}. We have only one parameter that controls both the 
thickness of the wall and the $AdS$ curvature. The limit $c\to\infty$ 
sends the width of the wall to zero and the $AdS$ curvature to infinity. 
Clearly classical gravity is not adequate to describe the physics in this
limit, but as a consistency check we can still compare \eq{se} to the 
corresponding equation in \cite{RS}. We expect these equations to agree
in the high curvature limit. We find for $z\neq 0$
\begin{equation}
V_{QM}(z) = \frac{15}{4z^2}.
\end{equation}
This agrees with the large curvature limit of the
potential given in \cite{RS} for $z\neq 0$, and solving \eq{se} with this
potential yields the Bessel functions found in \cite{RS}. We cannot check
that the singular piece in the potential of \cite{RS} comes out correctly,
because that requires first taking the width of the wall to zero and then
sending the curvature of $AdS$ to infinity. Our solution only allows us to
take a special correlated limit, so we do not expect the singular part to
agree. Finally, in our setup modes with masses up to $m_{max} \sim c$ are
suppressed on the domain wall. $m_{max}$ becomes infinitely large in the limit
$c\to \infty$,
so we recover uncorrected four-dimensional gravity. Correspondingly, 
we find that the corrections from bulk modes in the thin wall solution of
\cite{RS} vanish as the $AdS$ curvature goes to infinity.  

\acknowledgements 
It is a pleasure to thank Josh Erlich, Yael Shadmi, Yuri Shirman, and
especially Lisa Randall for helpful comments and for encouraging
me to write this paper. I would also like to thank the ITP at UCSB for
hospitality while this work was completed.
This work was supported in part by DOE grants \#DF-FC02-94ER40818 and
\#DE-FC-02-91ER40671 and NSF grant PHY94-07194.

{\tighten

} 

\begin{references} 

\bibitem{RS}
L.~Randall and R.~Sundrum,
``An alternative to compactification,''
hep-th/9906064.
%%CITATION = HEP-TH 9906064;%%

\bibitem{review}
M.~Cvetic and H.H.~Soleng,
``Supergravity domain walls,''
Phys.\ Rept.\ {\bf 282}, 159 (1997)
hep-th/9604090.
%%CITATION = PRPLC,282,159;%%

\bibitem{mark}
W.D.~Goldberger and M.B.~Wise,
``Modulus stabilization with bulk fields,''
hep-ph/9907447.
%%CITATION = HEP-PH 9907447;%%

\bibitem{gubs}
O.~DeWolfe, D.Z.~Freedman, S.S.~Gubser and A.~Karch,
``Modeling the fifth dimension with scalars and gravity,''
hep-th/9909134.
%%CITATION = HEP-TH 9909134;%%

\bibitem{cosm}
C.~Csaki, M.~Graesser, C.~Kolda and J.~Terning,
``Cosmology of one extra dimension with localized gravity,''
hep-ph/9906513.
\\
H.B.~Kim and H.D.~Kim,
``Inflation and gauge hierarchy in Randall-Sundrum compactification,''
hep-th/9909053.
%%CITATION = HEP-TH 9909053;%%
\\
P.~Kanti, I.I.~Kogan, K.A.~Olive and M.~Pospelov,
``Cosmological 3-brane solutions,''
hep-ph/9909481.
%%CITATION = HEP-PH 9909481;%%
\\
P.~Kraus,
``Dynamics of anti-de Sitter domain walls,''
hep-th/9910149.
%%CITATION = HEP-TH 9910149;%%
\\
A.~Kehagias and E.~Kiritsis,
``Mirage cosmology,''
hep-th/9910174.
%%CITATION = HEP-TH 9910174;%%
\\
T.~Shiromizu, K.~Maeda and M.~Sasaki,
``The Einstein equation on the 3-brane world,''
gr-qc/9910076.
%%CITATION = GR-QC 9910076;%%
\\
%%CITATION = HEP-PH 9906513;%%
P.~Binetruy, C.~Deffayet, U.~Ellwanger and D.~Langlois,
``Brane cosmological evolution in a bulk with cosmological constant,''
hep-th/9910219.
%%CITATION = HEP-TH 9910219;%%
\\
E.E.~Flanagan, S.H.~Tye and I.~Wasserman,
``Cosmological expansion in the Randall-Sundrum brane world scenario,''
hep-ph/9910498.
%%CITATION = HEP-PH 9910498;%%
\\
C.~Csaki, M.~Graesser, L.~Randall and J.~Terning,
``Cosmology of brane models with radion stabilization,''
hep-ph/9911406.
%%CITATION = HEP-PH 9911406;%%
\\
D.N.~Vollick,
``Cosmology on a three-brane,''
hep-th/9911181.
%%CITATION = HEP-TH 9911181;%%

\bibitem{csaki}
H.~Verlinde,
``Holography and compactification,''
hep-th/9906182.
%%CITATION = HEP-TH 9906182;%%
\\
N.~Arkani-Hamed, S.~Dimopoulos, G.~Dvali and N.~Kaloper,
``Infinitely large new dimensions,''
hep-th/9907209.
%%CITATION = HEP-TH 9907209;%%
\\
C.~Csaki and Y.~Shirman,
``Brane junctions in the Randall-Sundrum scenario,''
hep-th/9908186.
%%CITATION = HEP-TH 9908186;%%
\\
A.E.~Nelson,
``A new angle on intersecting branes in infinite extra dimensions,''
hep-th/9909001.
%%CITATION = HEP-TH 9909001;%%
\\
S.M.~Carroll, S.~Hellerman and M.~Trodden,
``BPS domain wall junctions in infinitely large extra dimensions,''
hep-th/9911083.
%%CITATION = HEP-TH 9911083;%%
\\
S.~Nam,
``Modeling a network of brane worlds,''
hep-th/9911104.
%%CITATION = HEP-TH 9911104;%%

\bibitem{bs}
A.~Brandhuber and K.~Sfetsos,
``Non-standard compactifications with mass gaps and Newton's law,''
JHEP {\bf 10}, 013 (1999) hep-th/9908116.
%%CITATION = JHEPA,9910,013;%%

\bibitem{sn}
S.~Nam, ``Mass gap in Kaluza-Klein spectrum in a network of brane worlds,''
hep-th/9911237.
%%CITATION = HEP-TH 9911237;%%

\bibitem{sdw}
K.~Behrndt and M.~Cvetic,
``Supersymmetric domain wall world from D = 5 simple gauged supergravity,''
hep-th/9909058.
%%CITATION = HEP-TH 9909058;%%
\\
A.~Chamblin and G.W.~Gibbons, ``Supergravity on the brane,'' hep-th/9909130.
%%CITATION = HEP-TH 9909130;%%
\\
D.~Youm, ``Solitons in brane worlds,'' hep-th/9911218.
%%CITATION = HEP-TH 9911218;%%

\bibitem{nogo}
R.~Kallosh, A.~Linde and M.~Shmakova,
``Supersymmetric multiple basin attractors,'' JHEP {\bf 11}, 010 (1999)
hep-th/9910021.
%%CITATION = JHEPA,9911,010;%%

\bibitem{fo1}
M.~Cvetic, S.~Griffies and S.~Rey, ``Static domain walls in N=1 supergravity,''
Nucl.\ Phys.\ {\bf B381}, 301 (1992) hep-th/9201007.
%%CITATION = NUPHA,B381,301;%%

\bibitem{fo2}
K.~Skenderis and P.K.~Townsend,
``Gravitational stability and renormalization-group flow,''
Phys.\ Lett.\ {\bf B468}, 46 (1999) hep-th/9909070.
%%CITATION = PHLTA,B468,46;%%


\end{references}
\end{document}